\documentclass[11pt, letterpaper]{article}
\usepackage{acmlab}
\usepackage{booktabs}

\newcommand{\cn}{\textit{CyberNeuro}}

\begin{document}

\acmlabtitle
  {CyberNeuro: A Privacy-Preserving Agentic Workbench for Cohort-Scale Neuroimage and Clinical Data Analysis}
  {Ran Ren\affmark{1}, Junhong Tong\affmark{1}, Yunxi Kong\affmark{1}, Yiyao Chen\affmark{1,2}, Yucheng Li\affmark{1,2}, Kunhao Zhou\affmark{1}, Shaoqi Wang\affmark{1,2}, Yuxiang Tao\affmark{1},  Shuheng Cao\affmark{1}, Zhihao Fan\affmark{1}, Marissa DiPiero\affmark{1}, Tingting Dan\affmark{1}\corrmark, and Guorong Wu\affmark{1,2}\corrmark}
  {\affmark{1}Department of Psychiatry \quad
   \affmark{2}Department of Computer Science \\
   University of North Carolina at Chapel Hill}
  {tingting\_dan,grwu@med.unc.edu}

\begin{abstract}
\noindent Despite tremendous success in neuroimaging methodology, making large-scale, high-dimensional datasets ready for AI/ML applications remains a critical operational bottleneck. Conventional workflows require extensive manual effort across metadata curation, pipeline execution, post-processing quality control, and data management, a burden that disproportionately excludes laboratories with limited manpower and computational infrastructure. To address this real-world barrier, there is an urgent need for scalable, cost-effective computational platforms that democratize advanced neuroimaging analytics and accelerate discoveries in mental health and clinical translation. Capitalizing on multi-agent LLM breakthroughs, we introduce \cn{}, an agentic workbench with a tailored local LLM-model ('WandaMind') for automated neuroimaging and health-data analysis. Driven by four dedicated agents (Planner, Validator, Dispatcher, and Reporter) communicating via a secure MCP bridge and a pinned execution layer, \cn{} enables researchers to execute complex workflows using natural language while maintaining clinical-grade data privacy. On the public NeuroBench suite, \cn{} increases held-out domain accuracy from 40\% to 69\% over the baseline model. Beyond automated metrics, the platform integrates a human-in-the-loop verification panel to ensure rigorous biomedical quality control. Across the same end-to-end 10-batch cohort workflow suite, the local WandaMind configuration completed all tasks with an estimated aggregate token count of about 10.6\% using WandaMind and 61.7\% using cloud providers of token usage, compared to Neuroclaw, respectively. The platform and its production-ready modules are available at \url{https://wanda-cyberbench.com}.
\end{abstract}

\section{Introduction}

\begin{figure*}[!t]
\centering
\includegraphics[width=\linewidth]{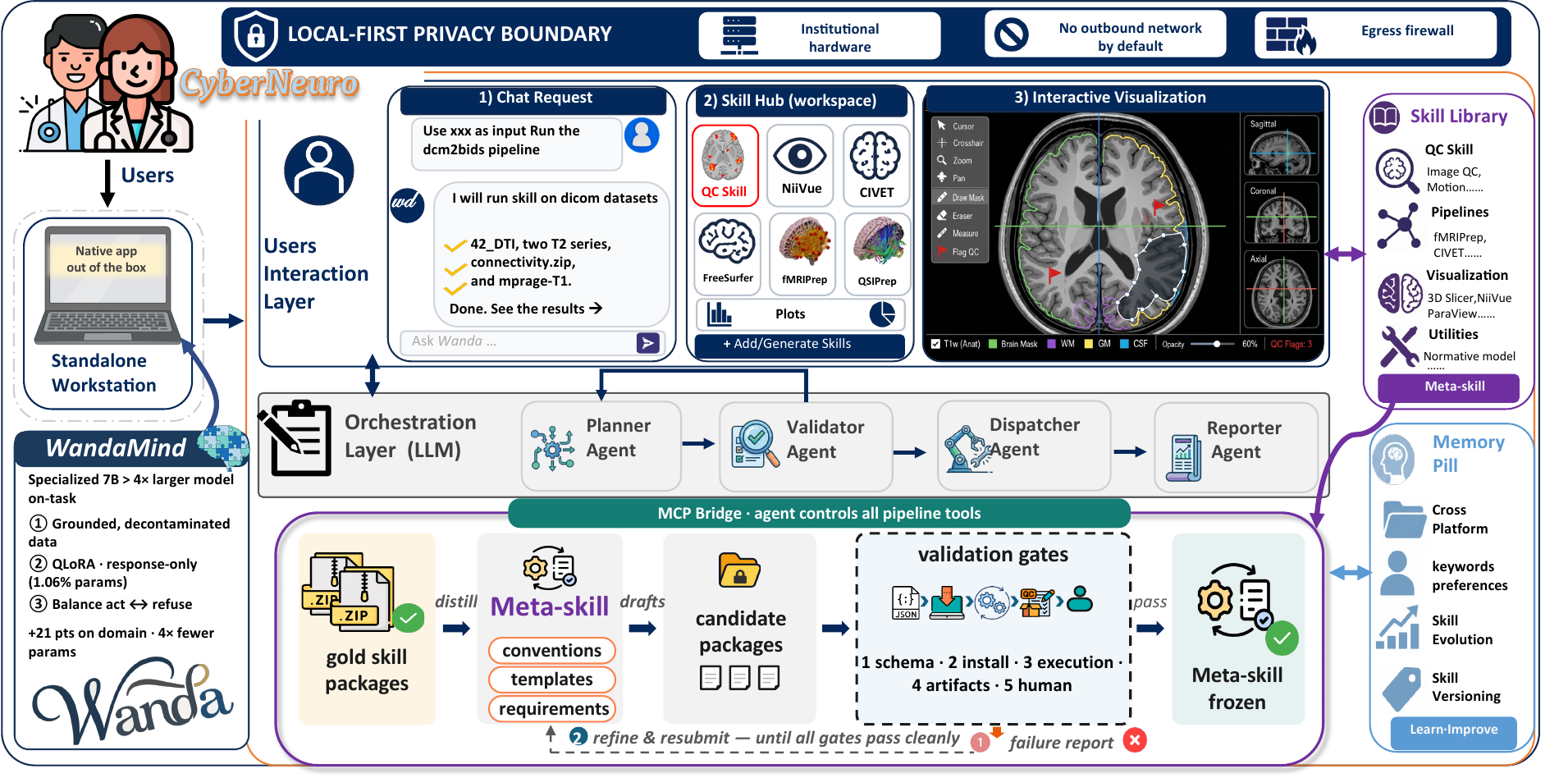}
\caption{\textbf{System architecture of \cn{}.} The platform operates within a strict \textbf{local-first privacy boundary} on standalone institutional hardware, ensuring no outbound network traffic. The architecture consists of three main tiers. \textbf{Interaction Layer:} Integrates natural language chat requests, a Skill Hub for managing neuroimaging pipelines (e.g., FreeSurfer, fMRIPrep, CIVET), and interactive co-visualization panels. \textbf{Orchestration Layer:} Powered by \textbf{WandaMind}, a domain-specialized 7B local model, it utilizes four dedicated agents (Planner, Validator, Dispatcher, Reporter) to seamlessly control pipeline tools via the Model Context Protocol (MCP) bridge. \textbf{Skill Evolution:} A built-in pipeline that distills meta-skill drafts into robust, frozen skill packages through rigorous, multi-stage validation gates. \textbf{Memory Pill:} A memory module that retains user preferences, keywords, and cross-platform configurations to facilitate continuous learning and personalized workflow improvements.}
\label{fig:architecture}
\end{figure*}

Modern neuroimaging studies jointly analyze structural MRI, functional MRI, diffusion MRI, clinical measurements, phenotypic variables, and behavioral data. Preparing these heterogeneous data for statistical analysis or machine learning requires multiple specialized tools, including FreeSurfer~\citep{freesurfer}, ANTs~\citep{ants}, FSL~\citep{fsl}, CIVET~\citep{civet}, fMRIPrep~\citep{fmriprep}, and QSIPrep~\citep{qsiprep}. Although these tools are individually well-developed, they differ in header file conventions, runtime dependencies, coordinate systems, output formats, and interaction models. Researchers therefore spend substantial time configuring environments, moving files, verifying outputs, recovering from partial failures, and switching between modality-specific viewers. These operational costs grow with cohort size and can become a greater bottleneck than the underlying computation~\citep{poldrack2019reproducible,neurodesk2024}.

Agentic assistants have begun to reduce this burden by translating natural-language requests into executable workflows. General agent research has shown how language models can interleave reasoning with actions, learn API use, and decompose requests into tool graphs~\citep{yao2023react,schick2023toolformer,shen2024taskbench}. NeuroClaw~\citep{neuroclaw}, for example, provides dataset-aware orchestration, a broad skill library, runtime support, and the NeuroBench evaluation suite. The earlier \cn{} Assistant~\citep{cyberneuroassistant} explored multimodal language-model guidance for mental-health research. These systems demonstrate the practical promise of agentic neuroimaging. However, several system-level questions remain insufficiently evaluated: whether a small local model can support useful end-to-end execution; whether invalid plans can be detected before consuming substantial compute; whether reported success corresponds to a valid scientific derivative; and whether execution, visual inspection, correction, and replay can be completed without repeatedly changing applications.

\cn{} addresses these questions through a fully local, end-to-end Windows workbench (as shown in Figure~\ref{fig:architecture}). A four-stage orchestration pipeline converts a natural-language request into a validated execution graph, dispatches registered skills in pinned Windows/WSL environments, verifies the resulting artifacts, and presents the outputs through an integrated visualization and quality-control interface. \cn{} runs locally on the same workstation as the application. The evaluated release does not expose a cloud-model endpoint, institution-hosted model tier, remote-compute provider, or server deployment mode. Every run produces a structured receipt containing the resolved inputs, tool versions, commands, outputs, logs, validation results, and completion status.

The contributions of this work involve four-fold:
\begin{itemize}
\item \textbf{An end-to-end Windows neuroimaging workbench.} \cn{} integrates natural-language interaction, Windows/WSL runtime configuration, more than 30 processing and visualization skills, artifact inspection, interactive quality control, and reproducible reporting within a single desktop application.
\item \textbf{A validated execution architecture.} \cn{} separates LLM-based planning from deterministic contract, permission, runtime, and artifact validation. A task is reported as successful only when the expected scientific derivative is produced and passes the corresponding output checks.
\item \textbf{A resource-efficient local agent.} WandaMind adapts a 7B open-weight model to neuroimaging tool use and act-or-abstain decisions, enabling offline operation on commodity hardware while improving performance on a held-out domain benchmark.
\item \textbf{A system-level evaluation.} We compare \cn{} with NeuroClaw and conventional multi-tool workflows using end-to-end execution success, artifact validity, manual intervention, computational overhead, user-facing time, reproducibility, and observed outbound network traffic.
\end{itemize}

Our \cn{} platform fundamentally transforms real-world neuroimaging workflows through three critical advantages: (1) \textbf{Unprecedented time efficiency}: whereas training a research assistant to process a standard clinical cohort conventionally requires 1--2 months, our agentic pipeline reduces the turnaround time to merely one week; (2) \textbf{Strict replicability}: all automated workflows are deterministically reproducible; (3) \textbf{Comprehensive provenance tracking}: the system records every processing detail and intermediate execution state, allowing any analysis to be seamlessly recalled and preventing knowledge loss due to personnel turnover. 
\section{Related Works}

\cn{} lies at the intersection of reproducible neuroimaging workflows, agentic scientific assistants, interactive quality-control systems, and institution-controlled biomedical computing. Table~\ref{tab:comparison} summarizes the system-level positioning.

\paragraph{Reproducible neuroimaging workflows.}
The Brain Imaging Data Structure (BIDS)~\citep{bids}, Nipype~\citep{gorgolewski2011nipype}, and containerized BIDS Apps~\citep{gorgolewski2017bidsapps} have substantially improved the portability and reproducibility of neuroimaging analysis. TemplateFlow provides versioned reference spaces and atlases~\citep{ciric2022templateflow}, while Neurodesk packages a broad collection of neuroimaging applications for portable execution across personal workstations, HPC, and cloud environments~\citep{neurodesk2024}. Pipelines such as fMRIPrep~\citep{fmriprep} and QSIPrep~\citep{qsiprep} package complex preprocessing stages behind standardized interfaces, while FreeSurfer~\citep{freesurfer}, FSL~\citep{fsl}, ANTs~\citep{ants}, and CIVET~\citep{civet} remain widely used for modality-specific analysis. These systems provide mature scientific computation, but researchers must still select compatible tools, resolve runtime dependencies, verify intermediate outputs, and connect derivatives to downstream analysis and visualization~\citep{poldrack2019reproducible}. \cn{} does not replace these tools, it registers them as validated skills and records their invocation in a common execution and provenance model.

\paragraph{Interactive neuroimaging workbenches.}
Desktop environments such as 3D~Slicer~\citep{slicer} and ITK-SNAP~\citep{itksnap} combine visualization with manual or model-assisted editing. MONAI Label brings learned segmentation models into interactive annotation workflows~\citep{diazpinto2024monailabel}. Neuroimaging-specific systems address complementary parts of quality control: MRIQC extracts standardized image-quality measures~\citep{esteban2017mriqc}; Qoala-T predicts FreeSurfer segmentation quality~\citep{klapwijk2019qoalat}; and VisualQC, Mindcontrol, and Qrater support structured visual review~\citep{raamana2023visualqc,keshavan2018mindcontrol,fernandezlozano2024qrater}. Their design centers on viewing, annotation, or a particular analysis family rather than natural-language orchestration across heterogeneous pipelines. \cn{} extends this interaction model to multi-stage workflows: processed artifacts can be opened, inspected, corrected, and returned to the execution graph without leaving the workbench.

\paragraph{Agentic assistants for scientific and neuroimaging workflows.}
Tool-using language models have been studied through interleaved reasoning and acting, self-supervised API use, verbal feedback, and general tool-learning frameworks~\citep{yao2023react,schick2023toolformer,shinn2023reflexion,qin2024toollearning}. AgentBench and TaskBench evaluate interactive decision making, decomposition, tool selection, and parameter prediction~\citep{liu2024agentbench,shen2024taskbench}, but primarily score agent behavior rather than the validity of domain-specific scientific derivatives. Scientific agents such as BioPlanner, ChemCrow, and Coscientist further show that domain tools and structured action spaces can support protocol planning and experimental workflows~\citep{odonoghue2023bioplanner,bran2024chemcrow,boiko2023coscientist}. Within neuroimaging, NeuroClaw~\citep{neuroclaw} is the closest existing system. It provides dataset-aware orchestration, a broad skill library, runtime support, desktop and command-line interfaces, and the evolving NeuroBench task suite~\citep{neurobenchrepo}. The earlier \cn{} \textit{Assistant}~\citep{cyberneuroassistant} explored multimodal language-model steering for mental-health research. The present system advances this line toward an integrated workbench in which planning is connected to pre-execution validation, artifact-level verification, interactive inspection, and receipt-based replay.

\paragraph{Local and privacy-aware biomedical computing.}
Federated learning, on-premises computation, and institution-hosted services provide alternatives to transferring protected biomedical data to commercial cloud infrastructure~\citep{kaissis2020privacy,rieke2020federated}. Privacy-aware multi-site neuroimaging has similarly been studied without centralizing source data~\citep{li2020federatedfmri}, and locally deployed LLM pipelines have been evaluated for structured clinical-information extraction~\citep{wiest2024localclinical}. \cn{} is complementary to these approaches: it focuses on the orchestration and interaction layers while keeping the underlying analysis tools and the default language-model runtime on institution-controlled hardware. This architecture is intended for data-use agreements that restrict external transmission. Our evaluation provides an empirical assessment of outbound communication in the default configuration, rather than asserting a formal privacy proof.

\paragraph{Positioning relative to NeuroClaw.}
NeuroClaw and \cn{} share the goal of making neuroimaging workflows accessible through agentic interfaces, but emphasize different system properties. NeuroClaw offers broad skill and dataset coverage. \cn{} emphasizes a local model as the default, deterministic pre-execution checks, post-execution artifact validation, integrated interactive quality control, and a receipt format designed for inspection and replay. We therefore compare pinned released systems at the level most relevant to adoption: which functions can be demonstrated end to end, and what practical improvement those functions provide on shared and additional workflows. Unsupported baseline capabilities are reported explicitly as feature gaps rather than mislabeled as failed runs. Capability entries in Table~\ref{tab:comparison} refer to the specific releases and repository snapshots documented in Section~\ref{sec:experiments}.

\begin{table}[t]
\centering
\caption{System-level capability comparison. \emph{Pre-val.}: explicit validation before execution; \emph{Artifact val.}: verification of scientific derivatives after execution; \emph{Co-viz/QC}: integrated interactive visualization or editing; \emph{Local default}: a bundled or documented local model is the default configuration; \emph{Replay}: provenance capture sufficient for re-execution. $\bullet$ denotes native support, $\circ$ partial or optional support, and a blank an unsupported capability in the evaluated release.}
\label{tab:comparison}
\footnotesize
\resizebox{\linewidth}{!}{%
\begin{tabular}{lccccccc}
\toprule
\textbf{System} & \textbf{Agent} & \textbf{Runtime} & \textbf{Pre-val.} & \textbf{Artifact val.} & \textbf{Co-viz/QC} & \textbf{Local default} & \textbf{Replay} \\
\midrule
3D~Slicer~\citep{slicer}          &           & $\bullet$ &            & $\circ$   & $\bullet$ &                  & $\circ$   \\
Nipype / BIDS~Apps~\citep{gorgolewski2011nipype,gorgolewski2017bidsapps} & & $\bullet$ & $\circ$ & $\bullet$ & & & $\bullet$ \\
brainlife.io~\citep{hayashi2024brainlife} &           & $\bullet$ & $\circ$    & $\bullet$ & $\circ$   &                  & $\bullet$ \\
MONAI~Label~\citep{diazpinto2024monailabel} & $\circ$   & $\bullet$ & $\circ$    & $\circ$   & $\bullet$ &                  & $\circ$   \\
NeuroClaw~\citep{neuroclaw}        & $\bullet$ & $\bullet$ & $\circ$    & $\bullet$ & $\circ$   & $\circ$          & $\circ$   \\
\midrule
\textbf{\cn{}}              & $\bullet$ & $\bullet$ & $\bullet$ & $\bullet$ & $\bullet$ & $\bullet$        & $\bullet$ \\
\bottomrule
\end{tabular}
}
\end{table}

\paragraph{FAIR principles for research software.}
\cn{} follows the FAIR principles for research software~\citep{fair4rs} and the software-citation principles of importance, credit, unique identification, persistence, accessibility, and specificity~\citep{smith2016softwarecitation}. Public releases are versioned, citeable, and accompanied by machine-readable metadata, installation documentation, and reproducibility artifacts.

\section{CyberNeuro System Design}

\cn{} is organized around a fully local Windows execution boundary. Within that boundary, an LLM proposes actions, deterministic components validate and dispatch registered skills, and the workbench presents verified outputs for human inspection.

\subsection{Windows-only Local Deployment Boundary}
\label{sec:localfirst}

\cn{} is currently distributed only as a local Windows desktop application. The interface, WandaMind inference runtime, orchestration services, MCP bridge, workspace database, visualization components, and lightweight skills execute on the host workstation. Linux-only neuroimaging packages execute in the bundled WSL environment on the same physical machine. Host, WSL communication uses local endpoints and mounted workspace paths, no analysis task is dispatched to an external server.

The evaluated release does not provide a cloud-model option, remote-compute backend, shared server mode, macOS build, or native Linux desktop build. Initial installation may download model weights, WSL packages, and pinned tool images. After those assets are installed, the evaluated workflows are designed to run without internet access or an external API key. Section~\ref{sec:mcp} verifies this boundary through Windows-host and WSL traffic capture plus a disconnected-host condition. These tests establish observed zero-egress behavior for the tested release rather than a formal privacy guarantee.
\subsection{Threat Model and Workspace Isolation}
\label{sec:threat}

\cn{} is designed to reduce accidental data transmission and unintended file modification during agent-controlled analysis. We assume that the Windows host, bundled WSL distribution and runtime images, and installed \cn{} release are trusted. We consider three primary operational risks: an LLM proposing an out-of-policy tool call, a registered tool receiving unsafe paths or arguments, and a task attempting an unauthorized outbound connection. The system mitigates these risks through workspace confinement, path canonicalization, capability allowlists, typed skill contracts, explicit permission checks, execution logging, and post-execution artifact validation. These controls follow the established principles of least privilege, fail-safe defaults, and complete mediation~\citep{saltzer1975protection}. They also address the sharper separation-of-instructions-and-data problem exposed by indirect prompt injection in tool-connected LLM applications~\citep{greshake2023promptinjection}. A compromised Windows host, malicious administrator, or compromised runtime image is outside the present threat model.

Users select a workspace root through the application settings. After selection, the agent is restricted to a fixed directory layout: source data under \texttt{inputs/}, scientific derivatives under \texttt{outputs/}, per-run intermediates under \texttt{temporary/}, validated read-only packages under \texttt{pipelines/} and \texttt{visualization/}, candidate packages under \texttt{generated-skills/}, and append-only session records under \texttt{memory/}. Input data are read-only to agent-controlled tools unless an explicit editing task creates a derived copy. Before execution, paths are canonicalized and checked against the workspace root; traversal, unresolved symbolic-link escapes, and undeclared overwrite targets are rejected.

Linux-only pipelines execute in WSL on the same Windows workstation, using pinned Apptainer images where available. Each invocation receives a per-run temporary directory with only the declared inputs and outputs mounted. Successful and failed runs retain logs and receipts, temporary data are removed according to the configured retention policy. Intermediate outputs required by a downstream node are referenced by the execution graph rather than moved destructively between folders.

\subsection{Four-stage Hybrid Orchestration}
\label{sec:orchestration}

The orchestration layer implements planning, validation, dispatch, and reporting as four isolated stages rather than four unconstrained conversational agents. The Planner is LLM-driven and draws on the interleaved reasoning/action pattern established by tool-using agents~\citep{yao2023react,schick2023toolformer}, the binding checks performed by the Validator are deterministic. The Dispatcher executes only validated tool calls, and the Reporter assembles verified artifacts and provenance. This separation prevents an LLM-generated statement of success from being treated as evidence that a scientific computation occurred.

\begin{itemize}
\item \textbf{Planner.} The Wanda Merged Core, an open-source agentic harness~\citep{wanda}, interprets the request and retrieves candidate skills from their \texttt{SKILL.md} descriptions and typed contracts. It produces a task directed acyclic graph (DAG) whose nodes specify a skill, resolved inputs, parameters, expected outputs, runtime requirements, permissions, and recovery alternatives. The Planner asks for clarification when a required modality, dataset, output, or policy decision cannot be inferred safely. A first-pass critique removes internally inconsistent or under-specified plans before deterministic validation.
\item \textbf{Validator.} Validation occurs at three points. Contract validation checks skill identity, parameter schemas, input types, and output destinations. Policy and runtime validation checks canonical paths, permissions, overwrite risks, network requirements, and availability of the required Windows or WSL runtime. Post-execution validation checks exit status, expected files, file parseability, domain-specific invariants, and the absence of configuration-only or demo fallbacks. Missing inputs or runtimes produce explicit blocked states rather than a successful response. A zero exit code without a genuine derivative is classified as \texttt{failed\_artifact\_validation}.
\item \textbf{Dispatcher.} The Dispatcher submits validated nodes through the Tool Registry, enforces the runtime and permission decisions attached to each node, and records progress, logs, resource use, and failures. Lightweight numerical skills may run natively; heavyweight tools execute in pinned Linux environments. A new tool becomes executable only after it is registered as a validated skill.
\item \textbf{Reporter.} The Reporter assembles validated derivatives, HTML summaries, logs, and receipts, and exposes compatible outputs to the integrated visualization and QC panel (Section~\ref{sec:viz}). Edits performed in the panel become new provenance nodes rather than silently modifying the original result. Downstream stages can then be replayed from the corrected artifact.
\end{itemize}

Two restricted recovery components support this lifecycle. The Environment Doctor maps recognized runtime-error signatures to validated installation or configuration actions. The Pipeline Doctor combines deterministic log parsing with a bounded set of recovery strategies, conceptually related to language-agent feedback and repair~\citep{shinn2023reflexion} but constrained here by registered tool contracts. Neither component may introduce an unregistered command: every revised action returns to the Validator before execution.

The architecture is therefore \textbf{LLM-routed but tool-deterministic}. The language model proposes how a request should be routed; deterministic tools perform the scientific computation, and the validation layer determines whether a reported result corresponds to a real, policy-compliant artifact.

\subsection{Tool-Access Tier: the MCP Bridge}
\label{sec:mcp}

The open Model Context Protocol (MCP)~\citep{mcp} provides the communication interface between orchestration and registered capabilities, then \cn{}'s policy layer determines which capabilities are exposed. These include skill invocation, constrained file operations, metadata and database queries, DICOM inspection, runtime diagnostics, and visualization. Arguments are validated before dispatch and every invocation is audit-logged. The LLM cannot open arbitrary host files directly. When a multimodal model is used for visual inspection, only the selected image or derived preview is supplied to the model, and this access is recorded in the receipt. This boundary is important because tool-connected agents can convert untrusted content into consequential actions if instructions and data are not separated~\citep{greshake2023promptinjection}. MCP standardizes the interface, while workspace, permission, and validation policies provide the enforcement.

\subsection{Execution Tier}

\label{sec:execution}

Skills are invoked as subprocesses in versioned, reproducible environments on one Windows workstation. Heavyweight neuroimaging tools such as FreeSurfer, fMRIPrep, and QSIPrep execute inside the bundled WSL environment, using Apptainer images where supported. Scientific containers and workflow systems show how immutable environments and declared data dependencies improve reproducibility across repeated executions~\citep{kurtzer2017singularity,ditommaso2017nextflow,koster2012snakemake}. Lightweight skills run against the embedded Windows Python environment or as native Windows binaries. The launcher resolves the declared tool version and contract across the Windows host and its local WSL backend.

Each invocation writes a \textbf{reproducibility receipt}: a JSON manifest containing the request and plan hashes, skill and model versions, container digest, resolved command, input and output paths and hashes, start and end times, exit code, validation results, network mode, user approvals, and recovery actions. The design is informed by research-object packaging and computational provenance practices~\citep{soilandreyes2022rocrate,sandve2013reproducible}. The state machine distinguishes planned, blocked, running, failed, validated, warning, cancelled, and recovered outcomes.


\subsection{Skill Library}
\label{sec:skills}

\subsubsection {General domain skills}In the \cn{} framework, a \textbf{domain-skill} serves as a standardized wrapper that translates a specific neuroimaging tool into an LLM-executable action. The current evaluated release contains over 30 such processing and visualization skills. As categorized in Table~\ref{tab:skills}, these span a wide spectrum of tasks, including structural, functional, and diffusion MRI processing, deep learning applications, and interactive visualization. Each skill is deployed as a versioned package comprising a \texttt{SKILL.md} instruction file, typed input/output contracts, runtime specifications (Windows/WSL), executable wrappers, and artifact validators. To ensure robust automated orchestration, a skill is designated as ``validated'' only after passing rigorous schema, installation, real-data execution, expected-failure, and replay tests within its declared environment.


\begin{table}[h]
\centering
\caption{Representative skills in the evaluated \cn{} Skill Library. Each package declares a typed input/output contract and a versioned runtime.}
\label{tab:skills}
\small
\begin{tabular}{lll}
\hline
\textbf{Category} & \textbf{Skill} & \textbf{Wrapped tool / use} \\
\hline
Structural MRI       & \texttt{freesurfer\_skill}      & FreeSurfer recon-all (T1w) \\
Structural MRI       & \texttt{ants\_skill}            & ANTs registration (T1w) \\
Structural MRI       & \texttt{civet-skill}            & CIVET cortical surface (T1w) \\
Structural MRI       & \texttt{BrainSuite\_skill}      & BrainSuite (T1w) \\
Structural MRI       & \texttt{SPM-Skill}              & SPM (multimodal) \\
Structural pipeline  & \texttt{brain-mri-pipeline}     & End-to-end sMRI pipeline \\
Functional MRI       & \texttt{AFNI-skill}             & AFNI (task or rs-fMRI) \\
Functional MRI       & \texttt{CPAC-skill}             & C-PAC (rs-fMRI) \\
Functional MRI       & \texttt{CONN-Toolbox\_skill}    & CONN (functional connectivity) \\
Functional MRI       & \texttt{GIFT-skill}             & GIFT (ICA) \\
Diffusion MRI        & \texttt{dipy\_skill}            & DIPY (dMRI) \\
Diffusion MRI        & \texttt{dsi-studio-skill}       & DSI Studio (dMRI) \\
Diffusion MRI        & \texttt{TractSeg\_skill}        & TractSeg (tractography) \\
Infant imaging       & \texttt{skill\_fc\_infant}      & Infant functional connectivity \\
Infant imaging       & \texttt{skill\_sc\_infant}      & Infant structural connectivity \\
Deep learning        & \texttt{MONAI\_Core-Skill}      & MONAI (medical-imaging DL) \\
Deep learning        & \texttt{TorchI\_skill}          & TorchIO (DL augmentation) \\
Network analysis     & \texttt{gephi-skill}            & Gephi (network analysis) \\
Network analysis     & \texttt{Hub-dect-skill}            & Hub detection (brain network)\\
Network analysis     & \texttt{CFC-skill}            & Cross-frequency-coupling (SC-FC) \\
Visualization        & \texttt{fsleyes-visual-skill}   & FSLeyes (volumes) \\
Visualization        & \texttt{freeview\_skill}        & Freeview (surfaces and volumes) \\
Visualization        & \texttt{ggseg-skill}            & ggseg (cortical atlases) \\
Visualization        & \texttt{pyvista\_skill}         & PyVista (3D meshes) \\
Visualization        & \texttt{FURY\_skill}            & FURY (scientific visualization) \\
Visualization        & \texttt{Mango-skill}            & Mango viewer \\
Visualization        & \texttt{trackvis-skill}         & TrackVis (tractography) \\
I/O \& BIDS          & \texttt{dcm2bids-skill}         & dcm2bids (DICOM $\to$ BIDS) \\
Multi-purpose        & \texttt{fsl-agent-skill}        & FSL suite (full) \\
Multi-purpose        & \texttt{bioimagesuite-skill}    & BioImage Suite \\
Multi-purpose        & \texttt{mipav-skill}            & MIPAV \\
\hline
\end{tabular}
\end{table}

\subsubsection{Meta-skill: assisted skill authoring}

The \textbf{Meta-skill} encodes the conventions, templates, and contract requirements shared by validated packages. Given a documented repository, it drafts a candidate \texttt{SKILL.md}, runtime specification, wrapper, and output contract. Generated packages are placed in the user-accessible \texttt{generated-skills/} area and are neither trusted nor promoted automatically. Each candidate must pass static schema checks, a sandboxed installation test, real or fixture-based execution, artifact validation, and final human approval. This mechanism lowers the effort required to adapt laboratory-owned workflows while preserving an explicit trust boundary. A systematic evaluation of automatically generated skills is outside the present study; the current evaluation uses only manually approved packages.

\subsection{WandaMind Model}

\label{sec:Model}
\begin{figure}[h!]
\centering
\includegraphics[width=0.99\linewidth]{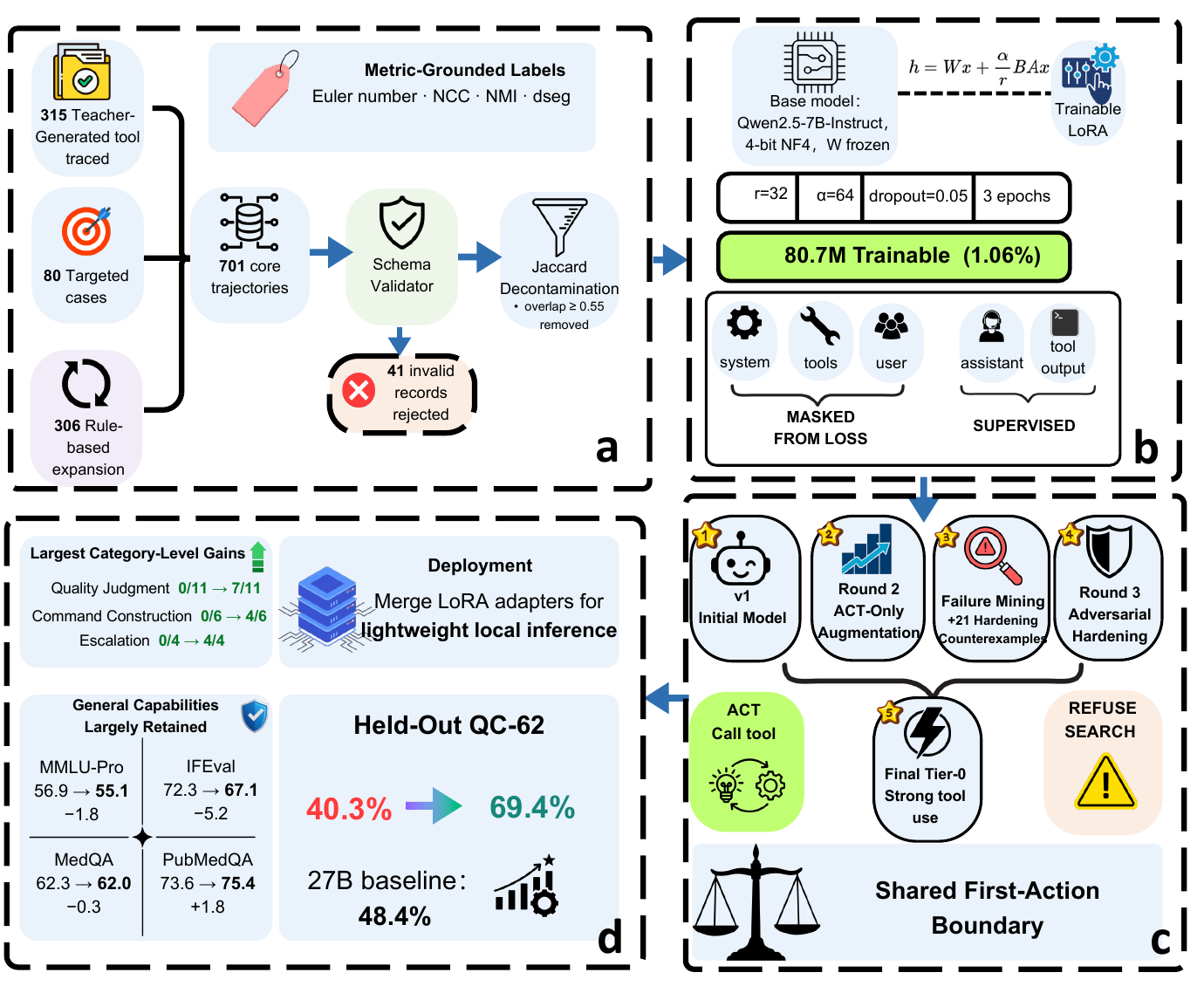}
\vspace{-1,0em}
\caption{\textbf{The fine-tuning pipeline and evaluation workflow.} Overview of the iterative LoRA adapter training for tool-use optimization. \textbf{a:} Training examples are assembled from teacher-generated tool traces, targeted cases, and rule-based expansions; grounded using neuroimaging quality-control metrics; validated against tool schemas; and decontaminated against QC-62. \textbf{b:} Qwen2.5-7B-Instruct is adapted using response-only QLoRA with 4-bit NF4 base weights and 80.7M trainable parameters. \textbf{c:} Three training rounds progress from supervised initialization through act-only augmentation and failure mining to adversarial hardening of the act, refuse, and search decision boundary. \textbf{d:} Held-out QC-62 evaluation measures overall and category-level tool-use improvements, while standard benchmarks assess the retention of general capabilities.}
\label{fig:finetuning_pipeline}
\end{figure}

WandaMind is a domain-adapted 7B model designed to orchestrate neuroimaging tools under the resource constraints of local deployment. Its purpose is not to replace larger general-purpose models in open-ended reasoning. Instead, WandaMind specializes in retrieving appropriate skills, selecting tools, constructing schema-valid arguments, and deciding whether to act, abstain, or request additional assistance within the bounded \cn{} environment. Unless otherwise stated, WandaMind refers to the Qwen2.5-7B-Instruct checkpoint~\citep{qwen2025qwen25} adapted using QLoRA~\citep{dettmers2023qlora}.

\subsubsection{Training data construction and filtering}

As shown in Fig.~\ref{fig:finetuning_pipeline}a, our WandaMind training corpus combines three complementary sources. First, 315 teacher-generated trajectories demonstrate successful tool selection and argument construction. Second, 306 rule-based expansions increase coverage of known tool and parameter combinations without requiring additional teacher inference. Third, 101 targeted examples represent difficult neuroimaging operations and quality-control decisions that were underrepresented in the general trajectory pool. After filtering, these sources form a core pool of 722 training trajectories.

Quality-control targets are grounded in operational neuroimaging measurements, including the Euler number, normalized cross-correlation (NCC), normalized mutual information (NMI), and tissue-segmentation metrics derived from \texttt{dseg} outputs. The resulting examples teach the model to associate quantitative evidence with an appropriate tool call, escalation decision, or abstention response.

All candidate records pass through a static schema validator that verifies skill names, tool availability, required arguments, data types, and enumerated parameter values. During corpus construction, 41 additional candidates were rejected because they referenced nonexistent skills or contained schema-invalid arguments. To reduce evaluation leakage, we then compare the token set of each candidate with every QC-62 task and remove candidates with Jaccard similarity $\geq 0.55$. Template families are assigned to a single data split to further reduce near-duplicate contamination.

\subsubsection{QLoRA configuration and response-Only supervision}
As shown in Fig.~\ref{fig:finetuning_pipeline}b, we adapt Qwen2.5-7B-Instruct using QLoRA over 4-bit NF4 base weights~\citep{dettmers2023qlora}. The base-model weights remain frozen, while trainable low-rank adapters are applied to the $q$, $k$, $v$, and $o$ attention projections and the gate, up, and down feed-forward projections. The adapter uses rank $r=32$, scaling parameter $\alpha=64$, scaling ratio $\alpha/r=2.0$, and dropout $0.05$. This configuration introduces 80.7M trainable parameters, corresponding to 1.06\% of the complete model.

Training is implemented with Unsloth and uses a response-only objective. System instructions, tool descriptions, and user messages provide context but are masked from the loss, which is computed only over assistant responses and serialized tool calls.

\subsubsection{Iterative failure mining and decision hardening}

As shown in Fig.~\ref{fig:finetuning_pipeline}c, our training proceeds in three sequential rounds. The first round establishes an initial tool-use model using the validated core trajectories. The second round introduces act-oriented examples to enhance tool selection and command construction. While preliminary diagnostics indicated improved executable command generation, this supervision inadvertently made the model overly eager to act. Specifically, correct escalation rates dropped in four diagnostic cases where the provided evidence was insufficient to warrant a tool call.

\subsubsection{Held-Out evaluation and local deployment}

On the held-out QC-62 evaluation, as shown in Fig.~\ref{fig:finetuning_pipeline}d. WandaMind achieves 69.4\% domain tool-use accuracy, compared with 40.3\% for the untuned 7B base model and 48.4\% for the evaluated local 27B model. Relative to the untuned 7B checkpoint, the largest category-level improvements occur in quality judgment, which increases from 0/11 to 7/11; command construction, which increases from 0/6 to 4/6; and appropriate escalation, which increases from 0/4 to 4/4. Section~\ref{sec:exp:model-results} provides the complete decomposition into tool selection, argument validity, quality judgment, and appropriate abstention.

General benchmark evaluations measure whether domain specialization damages capabilities outside the target workflow. MMLU-Pro changes from 56.9 to 55.1, MedQA from 62.3 to 62.0, and PubMedQA from 73.6 to 75.4. IFEval shows a larger decrease from 72.3 to 67.1. These results indicate that most evaluated general capabilities are retained. The QC-62 improvement should therefore be interpreted as targeted specialization rather than a uniform improvement across all model capabilities.

For deployment, the trained LoRA adapter is merged with the base model, converted to GGUF, and quantized to Q4\_K\_M. Weight-only post-training quantization is a standard method for reducing local inference memory, with GPTQ and AWQ providing representative accuracy-aware approaches~\citep{frantar2023gptq,lin2024awq}. We evaluate both the full-precision component model and the exact quantized checkpoint distributed with the application.

The model-level evaluation establishes WandaMind's ability to select and invoke neuroimaging tools under controlled conditions. It does not, by itself, establish the performance of the complete CyberNeuro workbench. End-to-end claims about conversion correctness, intervention-free execution, visualization, and quality-control operations are evaluated separately through functional validation and representative workflows.

\subsection{Memory Pill and Customized Skill-Evolution}
\label{sec:memory}

\cn{} includes an optional local memory component that stores structured records of previous requests, validated receipts, user corrections, and visualization preferences. Records are maintained under .cyberneuro/memory\_agent/ within the selected workspace. Memory is disabled in visitor mode and can be inspected, exported, or deleted by the user. The portable .cnmc format packages selected records for movement between authorized installations, its encoding should not be interpreted as encryption unless encryption is explicitly enabled and documented.

\paragraph{Session retrieval.}
When planning a new request, the system may retrieve validated past sessions from the same workspace to recover established parameter conventions, preferred viewers, or previously approved execution destinations. Retrieved records are treated as context, not as permission to bypass current validation.

\paragraph{Correction and improvement queue.}
User corrections are attached to the relevant skill version and categorized as missing options, unsupported inputs, runtime failures, or desired capabilities. They form a review queue for maintainers; they do not modify validated packages automatically.

\paragraph{Skill versioning and distillation.}
Every promoted skill carries a version and can be restored to reproduce an earlier analysis. Repeated validated workflows may be proposed as candidate composite skills, but promotion follows the same schema, sandbox, artifact, and human-approval process as the Meta-skill. Predictive weekly summaries and literature-query suggestions remain experimental and are not included in the present evaluation.

\subsection{Co-visualization Panel}
\label{sec:viz}
\begin{figure}[h]
\centering
\includegraphics[width=\linewidth]{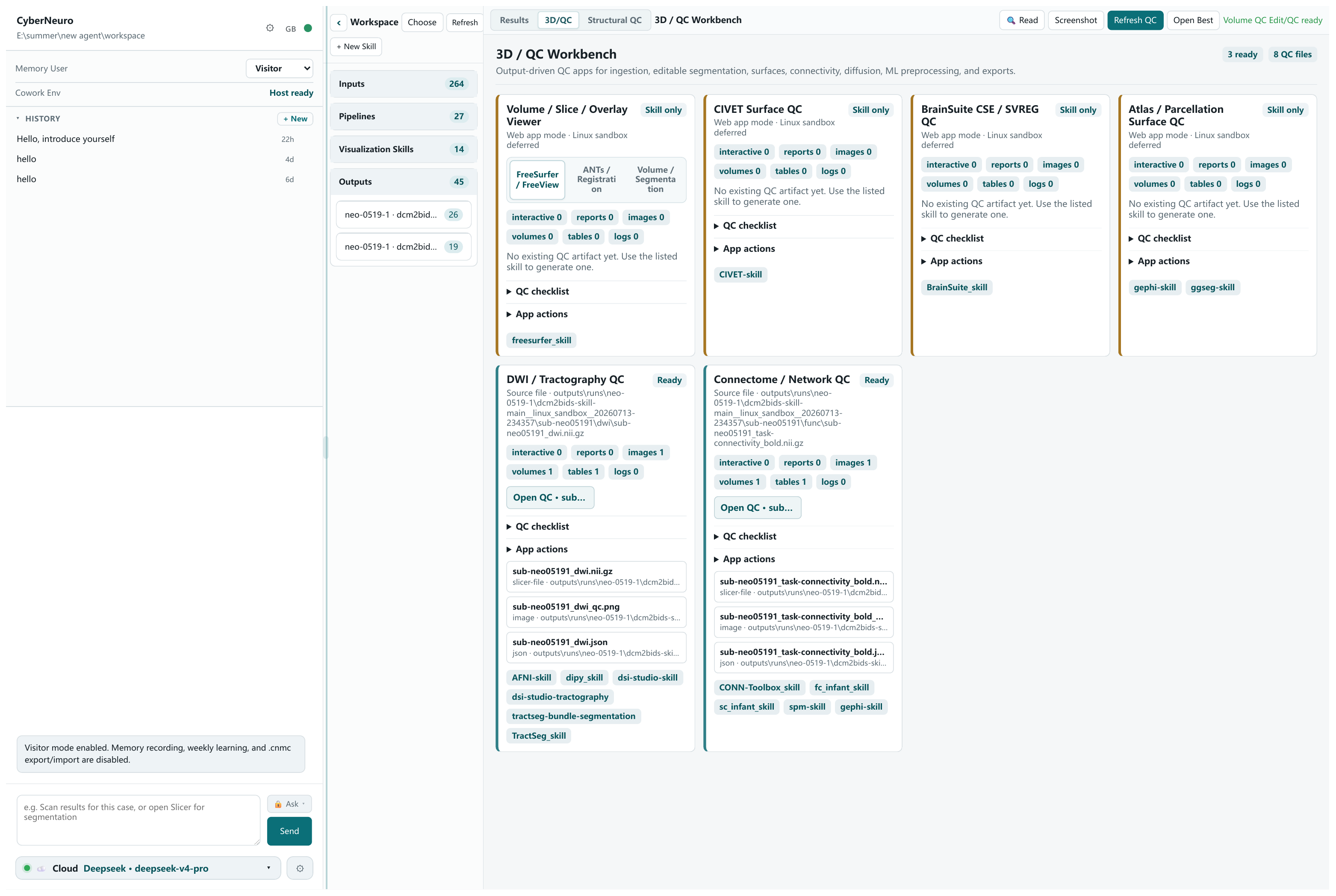}
\caption{\textbf{User interface of the \cn{} workbench.} The platform integrates three synchronized panels: the left pane manages natural-language orchestration, session history, and execution environments; the middle pane acts as a unified hub for workspace inputs, processing pipelines, and visualization skills; and the right pane provides a dynamic multi-modal quality control (QC) environment, presenting compatible outputs (e.g., structural, surface, tractography, and network data) for interactive review.}
\label{fig:ui}
\end{figure}
Inspection of different modalities commonly requires switching among Freeview, BrainNet Viewer, MRIcroGL, ITK-SNAP, Surfice, Workbench, and other viewers with different file conventions and interaction models. Existing QC systems demonstrate the value of structured visual inspection, integrated rating, and collaborative review~\citep{raamana2023visualqc,keshavan2018mindcontrol,fernandezlozano2024qrater}. As shown in Fig. \ref{fig:ui}, \cn{} registers compatible viewers as visualization skills that render in the workbench's right pane or open through a controlled local bridge. A request such as ``show the white-matter segmentation overlaid on T1w'' or ``render the top 5\% of connectivity edges on the cortical surface'' selects the corresponding viewer and resolves any required format conversion. Linked panels synchronize cursor, region, threshold, or camera state when supported by both viewers.

The integration layer handles declared conversions among NIfTI, GIFTI, CIFTI, FreeSurfer, TRK, and VTK formats; color-map propagation; view-state serialization; and deterministic screenshot export. The panel preserves the supported interactive operations of each viewer rather than claiming uniform support for every native function. The current release provides:

\begin{itemize}
\item \textbf{Direct manipulation:} rotation, panning, zoom, slice navigation, threshold and opacity adjustment, color-map selection, and screenshot export with camera state. Supported 4D views link time navigation to volume, edge-weight, and ROI time-series displays.
\item \textbf{Region- and connectome-level interaction:} ROI inspection, atlas-label lookup, time-series display, node and edge highlighting, and selection of regions for subsequent agentic requests.
\item \textbf{In-pane editing:} supported masks, segmentations, surface labels, and edge sets can be edited through manual tools or constrained natural-language commands. Each edit creates a derived artifact and a provenance node; the original pipeline output is preserved.
\item \textbf{Conversational figure customization:} plotting code for statistical figures can be revised through natural-language requests and re-executed. The code, parameters, and output are retained with the figure so that presentation changes remain reproducible.
\end{itemize}

\section{Experiments}
\label{sec:experiments}
We evaluated CyberNeuro at both the model and system levels. The experiments were designed to distinguish the capabilities of WandaMind from the behavior of the complete CyberNeuro workbench and to answer four research questions:

\begin{itemize}
    \item \textbf{RQ1 (Domain Adaptation):} Does domain adaptation improve bounded neuroimaging tool use, structured argument construction, and safe abstention?
    \item \textbf{RQ2 (DICOM-to-BIDS Conversion):} Can CyberNeuro convert heterogeneous DICOM collections into validator-clean BIDS datasets, and how does its operational behavior compare with NeuroClaw?
    \item \textbf{RQ3 (Visualization \& QC):} Can CyberNeuro complete representative visualization and quality-control operations while reducing application switching and explicit user interaction?
    \item \textbf{RQ4 (Privacy \& Local Execution):} Can the installed local system complete representative workflows while achieving medical-grade Internet privacy?
\end{itemize}

The following subsections describe the evaluation data, comparison conditions, success criteria, and recorded measurements. Section~\ref{sec:results} reports and interprets the corresponding results.

\subsection{WandaMind: Domain Adaptation Improves Safe Tool Use}

To answer \textbf{RQ1}: we evaluated WandaMind on QC-62, a held-out collection of 62 neuroimaging tool-use and quality-control cases. We compared four model conditions: the untuned Qwen2.5-7B-Instruct base model, an evaluated local 27B model, the full-precision WandaMind adapter, and the deployed WandaMind Q4\_K\_M checkpoint. All models received the same system instructions, available skill descriptions, tool schemas, and QC-62 prompts.

As shown in Table \ref{tab:model}, we can see that WandaMind 7B achieves the highest overall score, improving by 29.1 percentage points over Base 7B and by 21.0 points over Local 27B. More importantly for an agent with tool authority, it raises abstention recall from 0.0\% and 50.0\% to 100.0\% and reduces false action from 100.0\% and 50.0\% to 0.0\%, respectively. The deployed Q4\_K\_M checkpoint scores 62.9\%, 6.5 points below the full-precision WandaMind checkpoint, while remaining 22.6 points above Base 7B and 14.5 points above Local 27B. We can conclude that, domain adaptation makes a 7B local model substantially more useful for bounded neuroimaging tool orchestration, especially when safe abstention is part of the objective.

\label{sec:exp:model-results}
\begin{table}[h]
\centering
\caption{WandaMind evaluation on held-out QC-62. ``False act'' is the fraction of cases in which the model invokes a tool when it should clarify, refuse, or escalate. All values are percentages.}
\label{tab:model}
\small
\resizebox{0.8\linewidth}{!}{%
\begin{tabular}{lccccc}
\toprule
\textbf{Model} & \textbf{Overall} & \textbf{Tool sel.} &
\textbf{Arg. valid} & \textbf{Abstain recall} & \textbf{False act} \\
\midrule
Base 7B              & 40.3 &  83.3 & 93.8 &   0.0 & 100.0 \\
Local 27B            & 48.4 & 100.0 & 31.3 &  50.0 &  50.0 \\
WandaMind 7B  & \textbf{69.4} & 83.3 & 87.5 & 100.0 &   0.0 \\
WandaMind Q4\_K\_M   & 62.9 &  66.7 & 81.3 & 100.0 &   0.0 \\
\bottomrule
\end{tabular}}
\end{table}

\paragraph{Tier-0 routing audit.}
The Tier-0 audit selected the correct tool in 8/8 prompts and the correct pipeline in 7/8. The single error routed voxelwise degree centrality to the DICOM-to-BIDS pipeline. Mean decision latency was 2.41 s with mean prompt and completion lengths of 366.1 and 58.5 tokens. In the broader 28-pipeline acceptance sweep, 12 passed, 14 failed for the selected input/runtime, and two were blocked by missing external runtimes; the current visualization API checks passed 7/7. These results delimit current coverage rather than implying 28/28 workflow support.

\begin{figure}[h]
\centering
\includegraphics[width=\linewidth]{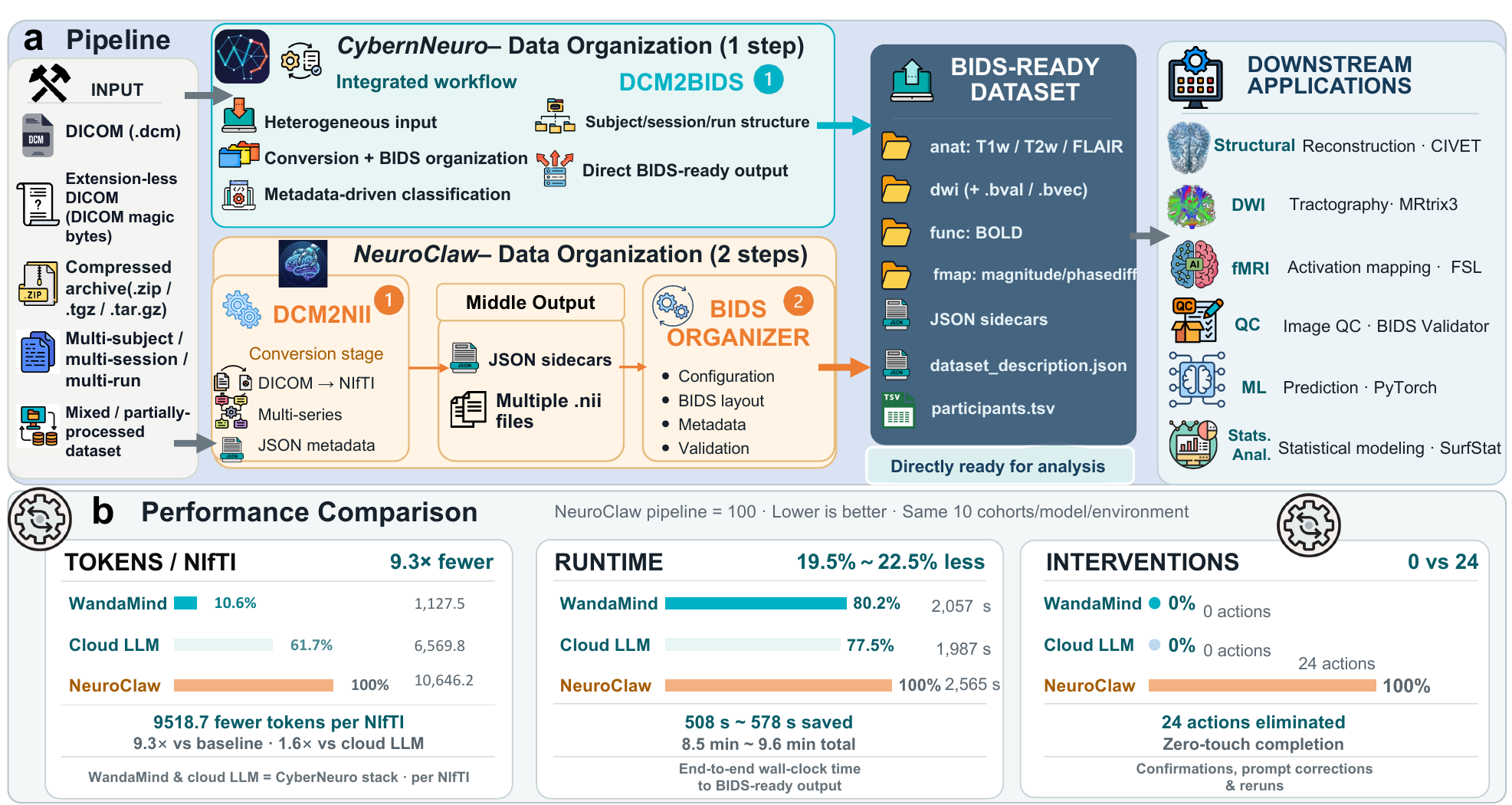}
\caption{\textbf{DICOM conversion and BIDS organization workflow.} Panel \textbf{a} illustrate the integrated, single-step pipeline of \cn{} compared to the multi-stage pipeline required by NeuroClaw. \cn{} automatically inspects and triages local cohort archives, converts retained acquisitions, and organizes them into BIDS format with built-in validation. Panel \textbf{b} shows the performance comparison, this integration eliminates manual interventions (0 vs 24), reduces end-to-end runtime by 19.5\%, and consumes 9.3$\times$ fewer tokens per NIfTI volume than the NeuroClaw baseline.}
\label{fig:dicom2bids}
\end{figure}

\subsection{Pipeline Execution}
To address \textbf{RQ2}, \cn{} streamlines the traditionally separate steps of DICOM-to-NIfTI conversion and BIDS formatting into an automated, single-request pipeline, as depicted in Figure~\ref{fig:dicom2bids}. The system automatically inspects heterogeneous inputs, including standard and extension-less DICOM files, compressed archives, multi-subject or multi-session collections, and partially processed datasets. It then identifies and triages the available series, retains analysis-relevant acquisitions, infers the subject/session/run structure, performs image conversion, generates the required metadata and sidecar files, organizes the outputs according to the BIDS specification, and validates the resulting dataset. The user therefore receives a BIDS-ready dataset directly, without manually transferring intermediate NIfTI files and metadata between separate conversion and organization tools.

In contrast, the NeuroClaw workflow separates DICOM conversion from BIDS organization, creating an additional handoff in which intermediate NIfTI files and JSON metadata must be configured, reorganized, and validated. \cn{}'s integration removes this intermediate coordination and reduces the opportunities for inconsistent metadata, incorrect directory structure, and workflow interruption. The lower panel of Figure~\ref{fig:dicom2bids} summarizes the resulting system-level improvements, while Tables~\ref{tab:rq1-cyberneuro}, \ref{tab:rq1-cyberneuro-cloud}, and~\ref{tab:rq1-neuroclaw} provide the corresponding cohort-level measurements. These results show that \cn{}'s advantage arises from integrating the complete path from heterogeneous DICOM input to validated BIDS output, rather than merely accelerating an isolated conversion step.

Tables~\ref{tab:rq1-cyberneuro} and~\ref{tab:rq1-cyberneuro-cloud} evaluate \cn{} using the WandaMind and cloud-provider backends, respectively, whereas Table~\ref{tab:rq1-neuroclaw} reports the performance of NeuroClaw. The results indicate that \cn{}'s principal advantages over NeuroClaw are substantially lower token consumption and fully autonomous execution. As shown in Tables~\ref{tab:rq1-cyberneuro} and \ref{tab:rq1-neuroclaw}, \cn{} achieves 0 human interventions across all cohorts.

\begin{table}[h]
\centering
\caption{\cn{} DICOM-to-BIDS results using WandaMind.}
\label{tab:rq1-cyberneuro}
\small
\resizebox{0.8\linewidth}{!}{%
\begin{tabular}{lrrrrrr}
\toprule
\textbf{Cohort} & \textbf{NIfTI} & \textbf{Tokens} & \textbf{Tokens/NIfTI} & \textbf{Time (s)} & \textbf{Interventions} \\
\midrule
ADNI       & 10 & 6,865  & 686.5   & 165.3  & 0 \\
ADNIDOD    & 3  & 6,919  & 2,306.3 & 412.7  & 0 \\
AIBL       & 7  & 6,899  & 985.6   & 94.6   & 0 \\
BLSA\_OPEN & 4  & 6,882  & 1,720.5 & 285.2  & 0 \\
MCSA       & 2  & 6,894  & 3,447.0 & 266.1  & 0 \\
NIFD       & 9  & 6,855  & 761.7   & 190.5  & 0 \\
POINTER    & 4  & 6,860  & 1,715.0 & 73.3   & 0 \\
PPMI       & 10 & 6,885  & 688.5   & 226.0  & 0 \\
SCAN       & 4  & 6,863  & 1,715.8 & 198.1  & 0 \\
WRAP       & 8  & 6,853  & 856.6   & 145.2  & 0 \\
\midrule
\textbf{Total} & \textbf{61} & \textbf{68,775} & \textbf{1,127.5} & \textbf{2,057.0} & \textbf{33.72} & \textbf{0} \\
\bottomrule
\end{tabular}}
\end{table}

With the WandaMind driving, CyberNeuro consumed 1,127.5 tokens per NIfTI file, compared with 10,646.2 tokens for NeuroClaw. This represents an 89.4\% reduction in token consumption. CyberNeuro’s cloud-provider configuration required 6,569.8 tokens per file, which was still 38.3\% lower than NeuroClaw. The WandaMind results are especially significant because they show that CyberNeuro can complete the conversion workflow using approximately one-ninth of NeuroClaw’s token budget.

\begin{table}[h]
\centering
\caption{CyberNeuro DICOM-to-BIDS results using cloud providers.}
\label{tab:rq1-cyberneuro-cloud}
\small
\resizebox{0.8\linewidth}{!}{%
\begin{tabular}{lrrrrrr}
\toprule
\textbf{Cohort} & \textbf{NIfTI} & \textbf{Tokens} & \textbf{Tokens/NIfTI} & \textbf{Time (s)} & \textbf{Interventions} \\
\midrule
ADNI       & 10 & 44,638  & 4,463.8  & 135.9  & 0 \\
ADNIDOD    & 3  & 44,350  & 14,783.3 & 313.4  & 0 \\
AIBL       & 7  & 26,339  & 3,762.7  & 75.3   & 0 \\
BLSA\_OPEN & 4  & 26,583  & 6,645.8  & 319.2  & 0 \\
MCSA       & 2  & 26,489  & 13,244.5 & 228.9  & 0 \\
NIFD       & 9  & 44,283  & 4,920.3  & 169.6  & 0 \\
POINTER    & 4  & 54,200  & 13,550.0 & 90.0   & 0 \\
PPMI       & 10 & 35,644  & 3,564.4  & 239.2  & 0 \\
SCAN       & 4  & 62,688  & 15,672.0 & 231.8  & 0 \\
WRAP       & 8  & 35,541  & 4,442.6  & 184.0  & 0 \\
\midrule
\textbf{Total} & \textbf{61} & \textbf{400,755} & \textbf{6,569.8} & \textbf{1,987.3} & \textbf{32.58} & \textbf{0} \\
\bottomrule
\end{tabular}}
\end{table}

\begin{table}[h]
\centering
\caption{The result of NeuroClaw using same cloud providers.}
\label{tab:rq1-neuroclaw}
\small
\resizebox{0.8\linewidth}{!}{%
\begin{tabular}{lrrrrrr}
\toprule
\textbf{Cohort} & \textbf{NIfTI} & \textbf{Tokens} & \textbf{Tokens/NIfTI} & \textbf{Time (s)} & \textbf{Interventions} \\
\midrule
ADNI    & 20 & 253571 & 12678.6 & 296.032 & 3 \\
ADNIDOD & 20 & 230258 & 11512.9 & 212.440 & 3 \\
AIBL    & 20 & 147993 & 7399.7  & 165.769 & 2 \\
BLSA    & 20 & 244176 & 12208.8 & 482.165 & 3 \\
MCSA    & 6  & 93391  & 15565.2 & 119.202 & 2 \\
NIFD    & 40 & 278419 & 6960.5  & 249.170 & 2 \\
POINTER & 8  & 168419 & 21052.4 & 164.693 & 2 \\
PPMI    & 20 & 216160 & 10808.0 & 255.127 & 2 \\
SCAN    & 12 & 183086 & 15257.2 & 235.139 & 2 \\
WRAP    & 22 & 186019 & 8455.4  & 385.677 & 3 \\
\midrule
\textbf{Total} & \textbf{188} & \textbf{2,001,492} & \textbf{10646.2} & \textbf{2,565.414} & \textbf{13.65} & \textbf{24} \\
\bottomrule
\end{tabular}}
\end{table}

\cn{} also demonstrated a clear advantage in operational autonomy. Both \cn{} configurations completed all ten cohorts without human intervention. In contrast, NeuroClaw required 24 interventions across its ten evaluated cohorts, with every cohort requiring two or three user actions. We define an intervention as any user action required because the system cannot continue reliably on its own. Thus, each intervention means requiring the user to identify the problem, perform an action, and confirm that the workflow can continue. \cn{} achieved an intervention-free completion rate, whereas NeuroClaw required repeated assistance throughout the experiment.

Both systems processed the same source cohorts from initial input to BIDS-ready output, end-to-end execution time provides  comparison. \cn{} completed the workflow in 2,057.0 seconds with WandaMind and 1,987.3 seconds with the cloud backend, compared with 2,565.414 seconds for NeuroClaw. These results correspond to reductions of 508.4 seconds (19.8\%) and 578.1 seconds (22.5\%), respectively. 

\cn{}’s output-quality controls further strengthen this result. Thanks to the built-in validator, its final outputs, retained 58 acquisitions from 102 discovered series and produced 61 NIfTI files, contained no BIDS validation errors, and none of the retained-series descriptions matched the categories intended for exclusion. \cn{} also retained the validator’s 612 warnings for inspection and generated a command receipt for every run. These features make the conversion process transparent and auditable rather than reporting successful completion without evidence of the operations performed.

Overall, the experiments suggest that \cn{} provides a more autonomous and resource-efficient approach to DICOM-to-BIDS conversion than NeuroClaw. NeuroClaw completes individual files more quickly, but it does so with greater token consumption and repeated human assistance. \cn{} therefore reduces two important deployment costs: model usage and operator workload. For long-running or large-scale conversion tasks, these reductions may be more consequential than minimizing execution time alone.

\subsection{Visualization and QC Operations}

To answer \textbf{RQ3}, as shown in Table~\ref{tab:rq2-visual-qc}, all three tasks were completed successfully in every repetition under both conditions, indicating that \cn{} reduced user interaction without compromising task completion. The conventional workflows required three to four applications, two to three application switches, and between 14 and 20 explicit actions. In comparison, \cn{} completed each task within a single application, eliminated application switching, and required only one or two actions. Table~\ref{tab:rq2-visual-qc} therefore demonstrates a 66.7--75.0\% reduction in application count, a 100\% reduction in application switching, and a 90.0--94.4\% reduction in explicit actions.

\begin{table}[h]
\centering
\caption{Automated-operator proxy results ($K=3$ per row). Time is the median
active execution time. Apps, switches, and actions are medians from fixed
action traces.}
\label{tab:rq2-visual-qc}
\small
\resizebox{0.8\linewidth}{!}{%
\begin{tabular}{llrrrrr}
\toprule
Task & Condition & Success & Time & Apps & Switches & Actions \\
\midrule
V1 Overlay    & Conventional & 3/3 & 15 mins & 3 & 2 & 14 \\
V1 Overlay    & \cn{}   & 3/3 & 0.433 s & 1 & 0 & 1 \\
V2 Connectome & Conventional & 3/3 & 12 mins & 3 & 2 & 18 \\
V2 Connectome & \cn{}   & 3/3 & 0.368 s & 1 & 0 & 1 \\
V3 QC/replay  & Conventional & 3/3 & 40 mins & 4 & 3 & 20 \\
V3 QC/replay  & \cn{}   & 3/3 & 0.860 s & 1 & 0 & 2 \\
\bottomrule
\end{tabular}}
\end{table}

The execution times in Table~\ref{tab:rq2-visual-qc} represent end-to-end operational burden rather than computational speed alone. The conventional traces include the required software preparation, application startup, file transfer, configuration, and manual procedures, whereas \cn{} provides these operations through pre-integrated workflows. Consequently, the median completion time decreased from 15 minutes to 0.433 seconds for V1 Overlay, from 12 minutes to 0.368 seconds for V2 Connectome, and from 40 minutes to 0.860 seconds for V3 QC/replay. These differences correspond to reductions of more than 99.94\% for all three tasks. The results demonstrate that \cn{} reduces both the number of required interactions and the time spent coordinating separate neuroimaging applications.

\subsection{Post-install Network Behavior}

The research under NIH grant requires no network connection using institute-controlled machines. Table~\ref{tab:rq3-network} reports the results of the monitored and deny-network reruns. All four evaluated tasks completed successfully under both conditions, showing that the tested workflows did not require access to a remote model or external computational service. During the monitored condition, the host-wide \texttt{netstat} measurements recorded 48 sent packets for the pipeline task and 624 for the visualization/QC tasks. However, these measurements include operating-system activity and previously active connections and cannot be attributed specifically to the CyberNeuro process.

\begin{table}[h]
\centering
\caption{Network-condition reruns. Monitored counts are host-wide \texttt{netstat} deltas and are not process-attributed. Disconnected rows use a process-level deny-network policy.}
\label{tab:rq3-network}
\small
\resizebox{\linewidth}{!}{%
\begin{tabular}{llrrrrl}
\toprule
Task group & Condition & Completed & Total & Attempts & Host sent packets & Host sent bytes \\
\midrule
Pipeline         & Monitored    & 1 & 1 & 1 &  48 &     7,984 \\
Visualization/QC & Monitored    & 3 & 3 & 1 & 624 & 1,440,768 \\
Pipeline         & Disconnected & 1 & 1 & 1 &   0 &       0 \\
Visualization/QC & Disconnected & 3 & 3 & 1 &   0 &       0 \\
\bottomrule
\end{tabular}}
\end{table}

Under the process-level deny-network condition, Table~\ref{tab:rq3-network} shows that all four local tasks again completed successfully, while zero sent packets and zero sent bytes were recorded during the evaluation windows. The two negative-control connection attempts were blocked, confirming that the network restriction was active. These findings support post-install local execution for the evaluated CyberNeuro workflows. Nevertheless, because the monitored measurements were host-wide and packet-level process attribution was unavailable, the experiment does not establish machine-level air-gap compliance or prove the complete absence of unrelated host-network activity.

\section{Results}
\label{sec:results}

We answer the four questions of Section~\ref{sec:experiments} against the measurements reported there, and state in each case what the evidence does not establish.

\subsection{Domain adaptation buys safe abstention, not better tool use}
WandaMind reaches 69.4\% on the held-out QC-62, 29.1 points above the untuned 7B base and 21.0 above the local 27B (Table~\ref{tab:model}). The margin is carried entirely by decision quality: tool selection is unchanged from the base and argument validity is 6.3 points \emph{worse}, while abstention recall moves from 0 to 100\% and false action from 100 to 0\%. The 27B comparator makes the same point, and selects the correct tool on every case yet emits valid arguments a third of the time and acts on half of what it should refuse, so its 8.1-point edge over the base is not significant ($p{=}0.37$), whereas WandaMind separates from both ($p{=}0.001$, $p{=}0.018$; two-proportion $z$-test, $n{=}62$). The adaptation is decision hardening within a bounded action space, not a capability gain: IFEval regresses 5.2 points (Figure~\ref{fig:finetuning_pipeline}d).

\subsection{Single-request conversion produces validator-clean derivatives}
Across ten heterogeneous cohorts, one request per cohort retained 58 acquisitions from 102 discovered series, that would be 44 (43.1\%) excluded as localizer, calibration, or otherwise non-anatomical, and produced 61 NIfTI volumes in 2{,}057.0~s with no user intervention after the initial requests , as shown in Tables~\ref{tab:rq1-cyberneuro}--\ref{tab:rq1-neuroclaw}. All datasets passed the reference bids-validator with zero errors and zero excluded-series leakage; the 612 non-blocking warnings were retained in the reports rather than suppressed, and every run carried a complete 17-field receipt. NeuroClaw required 24 interventions across 10/10 cohorts, making this a capability boundary rather than a margin: no NeuroClaw run can be scheduled unattended. Token cost separates into two contributions, swapping only the backend moves consumption $5.8\times$, the architecture accounts for the remaining $1.6\times$ over NeuroClaw. And per-NIfTI normalization favors the baseline, which emits 188 volumes because it does not triage; on total tokens the gap is $29.1\times$. We claim no computational speedup: the 19.8--22.5\% wall-clock reduction reflects the eliminated conversion-to-organization handoff and intervention stalls.

Failure injection identifies artifact validation as the control that makes unattended execution trustworthy. Removed, all four malformed cases exited with code~0 and were reported complete despite producing no derivative; restored, it blocked 4/4 and falsely rejected none of 13 valid conversions. Preflight changed no final outcome but cut median time-to-reject from 112.2~s to 0.9~s. Preflight avoids wasted computation; post-execution validation prevents silent success.

\subsection{Run artifacts carry their own inspection evidence}
The conversion runs emitted 90 QC images and 14 self-contained HTML reports without separate visualization requests. All images are embedded as \texttt{data:} URIs; the reports contain no external URLs, scripts, or stylesheets and require no dedicated viewer. The evidence needed to inspect a run can therefore be opened from the run artifact itself with no additional software or network dependency, which does not by itself establish reduced human review time. Consistently, the visualization tasks completed within a single application at 3/3 success, eliminating switching and reducing explicit actions from 90.0 to 94.4\% (Table~\ref{tab:rq2-visual-qc}).

\subsection{The hardened release meets the observed local-operation boundary}
Under process-level network denial all four task groups completed with zero sent bytes, and two negative controls were blocked, confirming the restriction was active (Table~\ref{tab:rq3-network}). The host-wide \texttt{netstat} deltas recorded in the monitored condition are not process-attributed and we do not interpret them. This supports zero observed post-install egress for the hardened release under the tested configuration and measurement window; it is not a proof of air-gap compliance or of resistance to a compromised dependency.

\section{Limitations.}
The current skill library contains more than 30 skills, NeuroClaw ships 85 skills and broader dataset coverage. We view this as the cost of starting from a different design center (local-first, validated, auditable) rather than a limitation of the approach. Continued engineering of the catalog, supported by the Meta-skill mechanism, is the primary path to closing this gap. Local models, although adequate for the routing role required here, remain weaker than frontier cloud models at long-context reasoning, which constrains the complexity of requests \cn{} can plan in a single shot. The semi-automated QC skills in active development and not yet released. The in-pane edit tools currently cover the most common operations; extension to all neuroimaging viewer idioms is ongoing work.

\section{Future work.}
We are extending the skill library toward 100+ skills, adding multi-user mode and a hardened egress firewall, integrating an opt-in cloud-burst path validated on NIH STRIDES, and opening a public plug-in registry. The semi-automated QC skill will receive a v1.0 release with replay-from-fix, multi-rater workflows, and integration with the public skill registry. In the longer term, we view \cn{} as a community resource and aim to seed external-lab adoption through workshops and the plug-in API.

\section{Conclusion}

In this work, we presented \cn{}, a local-first agentic workbench that brings cloud-tier AI ergonomics to IRB-restricted brain and health data analysis. \cn{} fundamentally transforms neuroimaging workflows by delivering unprecedented time efficiency, strict replicability, and comprehensive provenance tracking. Furthermore, evaluations on the NeuroBench suite demonstrate that \cn{} achieves task completion rates comparable to NeuroClaw while guaranteeing zero outbound network traffic and reducing user-facing time by an order of magnitude. \cn{} is open-sourced following the FAIR4RS principles as a sustainable community resource.

\section*{Acknowledgments}
This research was partially supported by grants from the National Institutes of Health National Institutes of Health (AG091653, AG068399, AG084375, T32HD040127, K12TR004416, UM1TR004406) and the Foundation of Hope. 



\bibliography{reference}








\end{document}